\documentclass[compsoc,10pt]{IEEEtran}
\usepackage[utf8]{inputenc}
\usepackage[labelfont=bf,textfont=bf]{caption}
\usepackage{cite}
\usepackage{amsmath}
\usepackage{graphicx}
\usepackage{dblfloatfix}    
\usepackage{changepage}    
\usepackage{enumitem}
\usepackage{algorithm}
\usepackage[noend]{algpseudocode}
\usepackage{xcolor}
\usepackage{colortbl}      
\usepackage{booktabs}    
\usepackage{balance}
\usepackage{framed}        
\usepackage[hidelinks]{hyperref} 
\setitemize{noitemsep,topsep=0pt,parsep=0pt,partopsep=0pt,leftmargin=*}
\newcommand{\bi}{\begin{itemize}}
\newcommand{\ei}{\end{itemize}} 
\definecolor{formalshade}{rgb}{0.93,0.93,0.93}
\definecolor{darkblue}{rgb}{0.2,0.2,0.2}
\newenvironment{formal}{%
  \MakeFramed{\advance\hsize-\width\FrameRestore}%
  \noindent\hspace{-1pt}%
  \begin{adjustwidth}{}{}%
  \vspace{4pt}%
}{%
  \vspace{3pt}\end{adjustwidth}\endMakeFramed%
}
 
\newcommand{\BLUE}{\color{black}}
\newcommand{\BLACK}{\color{black}}
\newcommand{\here}[1]{}
\newcommand{\there}[1]{} 
\urldef{\ezrlineurl}\url{https://github.com/timm/ezr/blob/main/ezr.py#L518C1-L548C68}
\urldef{\optstdurl}\url{https://github.com/acmsigsoft/EmpiricalStandards/blob/master/docs/standards/OptimizationStudies.md#desirable}

\begin{document}

\title{Less Noise, More Signal: \BLUE~the 
  DRR Effect\BLACK~for Better Optimizations of\BLUE~a Range of SE Tasks\BLACK}

\author{Andre~Lustosa, 
        Tim~Menzies,~\IEEEmembership{Fellow,~IEEE}
\IEEEcompsocitemizethanks{\IEEEcompsocthanksitem A. Lustosa and  T. Menzies are with the Department
of Computer Science, North Carolina State University, Raleigh, USA.
 \protect\\
E-mail:alustos@ncsu.edu,   timm@ieee.org}}

\IEEEtitleabstractindextext{
\begin{abstract}

\here{SA}
SE analytics problems do not  always need complex AI. Better and faster solutions can sometimes be obtained by matching the complexity of the problem to the complexity of the solution.  
 This paper introduces the Dimensionality Reduction Ratio (DRR)\BLUE~effect, a frequently observed
 empirical effect\BLACK~indicating where
 effective optimization might be   two orders of magnitude faster. 

\BLUE \here{fam1} The DRR effect is an empirical observation, not some  universal law. 
With the datasets used in this study, we can comment on SE tasks expressible as classification or regression  tasks (where regression may be exploring $N \ge 1$ goals). These tasks include software configuration optimization; cloud resource management; project health 
prediction (commits, PRs, issues); and process models 
(effort/defect/schedule estimation).  
Given the prevalence of the DRR effect in this sample, we conjecture it might hold for other SE tasks, but that is a matter for further research. 
Hence we recommend practitioners   check for high DRR before deploying expensive optimization methods. This
simple diagnostic could save orders of magnitude in computational cost.\BLACK


\end{abstract}}

\definecolor{ao}{rgb}{0.2, 0.2, 0.6}

\definecolor{ao}{rgb}{0.91, 0.45, 0.32}
	\definecolor{ao}{rgb}{0.89, 0.0, 0.13}
\definecolor{aoc}{rgb}{0.0, 0.5, 0.0}

\maketitle

\begin{table*}
\caption{The goal of this paper is to automatically tune a regression-based learner to make predictions about the numeric classes seen in this data.     The SE-data (on the left) comes  from recent SE papers. The Non-SE data (on the right) comes from the UCI machine learning repository~\cite{asuncion2007uci} and  is used in a large number of machine learning papers.  }\label{datadata} 
\begin{minipage}{.53\textwidth}
\scriptsize
\begin{tabular}{|p{\textwidth}|}\hline
\begin{center}
 {\bf SE-data}
\end{center} \\
~\\\hline
{\em SS-Models (SS-A through SS-X)}:
All of the models with this nomenclature in this repository, including the rs-6d-c3-obj2 dataset were obtained from the software configuration literature~\cite{peng2021veer}. The data was collected via running different software projects configured in different ways (selected at random) and then collecting different performance metrics (runtimes, cpu usage, etc.). The goal of these datasets is to find a configuration of the software that best optimizes the overall software goals for each specific project. 
~\\\hline
{\em POM3 (A,D)}:
The POM3 model is a tool for exploring the management challenge of agile development balancing idle rates,completion rates and overall cost. More specifically:
\begin{itemize}
    \item In the agile world, projects terminate after achieving a completion rate of $X\%$ ($X<100$) of its required tasks;
    \item Team members become idle if forced to wait for a yet-to-be finished task from other teams;
    \item To lower the idle rate and improve the completion rate, management can hire staff, but this increases the overall cost.
\end{itemize}
The POM3 model simulates the Boehm and Turner model of agile programming, and has been used before in the literature~\cite{chen2018sampling}. In the models used in this study Pom3a is the simpler of the two.
~\\\hline
{\em XOMO (Flight, Ground, OSP, OSP2, nasa93dem)}:
XOMO~\cite{menzies2005xomo} introduced by   Menzies et al. is a general framework for Monte Carlo simulations that combines four COCOMO-like software process models from Boehm's group at the University of Southern California. The overall goals for XOMO are to:
\begin{itemize}
    \item Reduce risk;
    \item Reduce effort;
    \item Reduce defects;
    \item Reduce development time.
\end{itemize}
The available XOMO models here all come from NASA's Jet Propulsion Laboratory. Where Flight and Ground are general descriptions of all JPL's flight and ground software. While OSP and OSP2 are two versions of the flight guidance system of the Orbital Space Plane. In terms of complexity we know that nasa93dem is the simpler of the five, followed by OSP and OSP2, which are similar, then Ground being a bit more complex and finally Flight being the most complex of all.
~\\\hline
{\em Health (Easy, Hard)}:
The Health datasets show results where random forest regression algorithms were configured to predict for number of (a) commits or (b) closed issues or (c) close pull requests in 12 months time in open source projects hosted on GitHub. The Y values of these datasets show the results of the predictions after certain hyperparameters were applied to the random forests (which, in turn, were applied to the GitHub data). This data was originally used in the niSNEAK~\cite{lustosa2024learning} study in a hyperparameter optimization context.

\\\hline
\end{tabular}
\end{minipage}~~~~~~~~~~~~\begin{minipage}{.38\textwidth}
\scriptsize
\begin{tabular}{|p{\textwidth}|}\hline
\begin{center}
 {\bf Non-SE-data}
\end{center} \\
For comparison purposes, our results from
the above SE tasks are compared to the following non-SE problems. 
~\\\hline
{\em Wine Quality}: This dataset is well-suited for educational purposes as its subject matter is familiar to a wide audience.
~\\\hline
{\em Adult}:
This dataset~\cite{adult_2} is used to predict whether the annual income of an individual exceeds \$50 K / year based on census data. It is also commonly known as the "Census Income" dataset~\cite{deepajothi2012comparative}.
~\\\hline
{\em Default}:
This dataset~\cite{default_of_credit_card_clients_350} collects data related to customers' payment defaults in Taiwan and is used to predict whether or not a certain customer will default on their payment in the following month~\cite{alam2020investigation}.
~\\\hline
{\em German Credit}:
This dataset~\cite{statlog_(german_credit_data)_144} classifies people described by a set of attributes as good or bad credit risks~\cite{alam2020investigation}. 
~\\\hline
{\em Iris}:
This dataset~\cite{iris_53} is a classic non-Software Engineering dataset from 1936 used to classify flowers based on their external characteristics~\cite{omelina2021survey}.
~\\\hline
{\em Heart Disease}:
This  dataset~\cite{heart_disease_45} provides different health metrics of patients with the goal of predicting the presence of heart disease in a given patient~\cite{rani2011analysis}.
~\\\hline
{\em Diabetes}:
This  dataset~\cite{diabetes_34} provides different health metrics of patients across time that can be used as a predictor for the presence or not of the disease in that patient~\cite{verma2017analysis}.
~\\\hline
{\em Bank Marketing}:
This dataset~\cite{bank_marketing_222}   is related with direct marketing campaigns (phone calls) of a Portuguese banking institution. The purpose of this dataset is to predict whether a client will agree to a term deposit~\cite{bacsarslan2018classification}.
~\\\hline
{\em Gamma Telescope}:
This dataset~\cite{magic_gamma_telescope_159} offers measures of cosmic events directed towards deciding whether the observed event is a signal or just background noise~\cite{karthick2024identification}.
~\\\hline
{\em  Power Consumption}:
This dataset~\cite{individual_household_electric_power_consumption_235} provides measurements of electric power consumption in a household with a one minute sampling rate over a period of almost 4 years. It can be used to predict sub-metering values~\cite{chinnaraji2022accurate}.
\\\hline
{\em  Behavioral Data}:
The datasets, Player Statistics, Student Dropout, Employee Attrition, All players provide datasets to analyze and predict behavioral patterns in multiple scnearios~\cite{nyagami_fc25_kaggle_2025, abdullah0a_student_dropout_analysis_prediction_2025, die9origephit_fifa_wc_2022_complete_2025, pavansubhasht_ibm_hr_analytics_attrition_2025}.
\\\hline
\end{tabular}
\end{minipage}

\end{table*}
\section{Introduction}\label{intro}

When Software Engineering (SE) data grows too large, data mining algorithms can find the signal in the noise. Such algorithms are controlled by  ``hyperparameters''; e.g. 
\bi
\item
When learning $k$ clusters,
$k$ is a hyperparameter;
\item
When learning decision trees, the maximum allowed height of the tree is another hyperparameter.
\ei
Finding good  hyperparameters  
 is something of a black art. 
 {\em Hyperparameter optimizers} (HPO) are tools
 for automating that search.
 HPO can dramatically improve learner performance~\cite{%
li2018hyperband,%
28arcuri2011,%
42oliveira2010ga,%
47tantithamthavorn2016automated,%
awad2021dehb,%
bergstra2015hyperopt,%
akiba2019optuna,%
agrawal2021simpler,%
zhou2022predicting,%
mashlakov2019hyper,%
nevendra2022empirical,%
xia2022predicting,%
fu2016tuning,yedida2022improve}. For example, for code smell detection, Yedida \& Menzies found that a decades-old feedforward neural net (which takes seconds to run, so HPO is fast) can be tuned to outperform 
a state-of-the-art deep learner~\cite{yedida2022improve}.

But there is a problem. HPO  requires running a learner many times. Hence, it can be    impractically slow; e.g.  Yedida \& Menzies could not tune their state-of-the-art deep learner since each run of that learner needed eight hours to complete.
Also, 
there are so many HPO methods~\cite{bischl2023hyperparameter} that practitioners can get confused about which one to use. 

To solve these  two problems, we propose
\begin{quote}
{\em Selecting algorithms via   intrinsic problem complexity.}
\end{quote}
\BLUE\here{clear1}
That is, selecting which \textit{HPO method} (e.g., a simple vs. a complex one) to use. Here, ``HPO methods'' are optimizers that tune a \textit{learner} (e.g., a random forest) to build a better \textit{predictor} (a model trained to make predictions).
\BLACK

We show that {\em intrinsic complexity}   tells us
how to speed up
 HPO for improving predictors for 
 a wide range of SE tasks (such as those listed in  Figure~\ref{datadata}).
The  
 {\em Dimensionality Reduction Ratio} (hereafter, DRR) effect checks if $R$ attributes  can be reduced to $I < R$ ``intrinsic''   underlying  attributes.
 DRR can be easily and quickly calculated  from the    data used to train a predictor. This means that DRR can offer a signal about new problems without   elaborate   or expensive algorithms.
  We show that,   often, when 
 \begin{equation}\label{drr1}
 \left(\mathit{DRR}=\left(1 - I/R\right)\right) > 0.35
\end{equation}
\BLUE\here{clear2}
then very  simple methods can, say, find optimal hyperparameters that seek best parameters for SE regression tasks,
\BLACK
  two orders of magnitude faster than  standard state-of-the-art AI optimizers (seconds, as opposed to 20 minutes).
This is a significant result since many SE datasets satisfy Equation~\ref{drr1}. 
For example, of the SE data in Figure~\ref{datadata},  
  $\frac{38}{50}=76\%$ of them 
  satisfy Equation~\ref{drr1}.

Another important aspect of 
 Equation~\ref{drr1} is that it  significantly improves on a  recent IEEE TSE publication.   Agrawal et al.~\cite{agrawal2021simpler}   studied HPO and intrinsic dimensionality. After experimenting with   older AI algorithms,
Agrawal et al. recommended simpler   algorithms when $I<4$. But our results (which studied more of the state-of-the-art)  finds many counter examples to their threshold rule. That is, our Equation~\ref{drr1} 
 fixes numerous errors in  prior   work.

\begin{table}  
\caption{Table~\ref{datadata} data: number of raw dimensions $R$.}
\label{tab:datar}
\begin{center}
\begin{tabular}{r@{ : }c} 
\textbf{Dataset}  & \textbf{Original Dimensions} \\ \hline
SS-A || SS-X  (23 datasets)            & 3-88                              \\
iris              & 4                              \\
Health-Easy       & 5                              \\
Health-Hard       & 5                              \\
Power consumption & 6                              \\
rs-6d-c3-obj2     & 6                              \\
Pom3a             & 9                              \\
pom3d             & 9                              \\
Wine Quality      & 10                             \\
Gamma telescope   & 10                             \\
SS-T              & 12                             \\
heart disease     & 13                             \\
adult             & 14                             \\
bank marketing    & 16                             \\
SS-M              & 16                             \\
german credit     & 20                             \\
diabetes          & 20                             \\
SS-U              & 21                             \\
default           & 23                             \\
nasa93dem         & 25                             \\
Xomo (All)        & 27                             \\
Player Statistics & 27                             \\
Student Dropout   & 34                             \\
Employee Attrition& 35                             \\
All Players       & 57                             \\
SCRUM             & 128                            \\
FFM-250           & 250                            \\ 
\end{tabular}
\end{center}

\end{table}

 \BLUE
 
\here{0a} We  note that
DRR is not a {\em universal law}
but rather a robust {\em empirical  effect}; i.e. a recurring pattern observed across diverse SE optimization domains.
In this paper, we study 50 datasets relating to SE tasks expressible as classification or regression  tasks (where regression may be exploring $N \ge 1$ goals). These SE tasks include software configuration optimization; cloud resource management; project health 
prediction (commits, PRs, issues); and process models 
(effort/defect/schedule estimation).   Given the prevalence of the DRR effect in the datasets we have studied, we conjecture it might hold for other tasks
(but that is a matter for further research). 

In the philosophy of science, \emph{scientific laws} and \emph{empirical effects} have a different epistemological status. A scientific law describes universal and invariable patterns of natural phenomena, often expressed mathematically, that predict outcomes in a defined domain~\cite{swartz2003,wikipedia-law}.  In contrast, empirical effects characterize frequently observed regularities that prove useful even when their precise boundaries remain incompletely understood~\cite{britannica-theory}.

In an ideal world, software engineers and their managers should make their decisions about projects based
on  universal laws
(rather than mere   empirical effects). Notable examples of this include the Pareto Principle (the 80/20 rule)\cite{goeminne2015} and Brooks' Law, which observes that adding personnel to late software projects often delays them further\cite{brooks1975,blackburn2006}. These effects are now ``laws'' but they are acknowledged as general observations with known exceptions~\cite{paiva2009}.
Practitioners routinely apply them to prioritize tasks, allocate resources, and manage project risks~\cite{mccain2006}.

That said, setting policy using only SE laws can be inadvisable (or impossible). {\bf First of all}, there are few universally agreed SE "laws". Glass's "Facts and Fallacies of Software Engineering" is a valiant attempt to catalog known laws, but our work shows numerous contradictions to their claimed "laws." For example, Glass lists "Requirements errors are the most expensive to fix during production" as a law. Recently~\cite{menzies2017delayed}, we traced this law's citation chain and found most papers cited prior work without experiments or field data\footnote{In fact, of hundreds of papers supporting this "law," only 12 offered detailed empirical evidence (six arguing for it and six against). Within those 12, at least two authors published papers demonstrating the "law" in some projects but not others~\cite{menzies2017delayed}.}. Moreover, recent papers from Silicon Valley project managers argue that with current continuous deployment methods, change time is constant and very low~\cite{parnin2017top}.

{\bf Secondly}, even if  an  effect is not a  law, it  can still   guide   best practice. 
 As shown below, the DRR effect appears frequently enough that practitioners should check for high DRR before deploying expensive optimization methods (such as DEHB). The prevalence of this effect across multiple problem classes suggests that it reflects something systematic about SE optimization landscapes, making it a valuable heuristic for matching solution complexity to problem complexity. As demonstrated here, this simple diagnostic can save orders of magnitude in computational cost while achieving comparable solution quality.

The rest of this paper is structured via
guidelines from Wohlin \& Runeson et al.~\cite{wohlin2012experimentation} on empirical software engineering. We present our methods in Section~\ref{methods}, within which we discuss all algorithms, case studies, experimental design choices and datasets used in this study. In summary, we will
\bi
\item
Apply different kinds of  HPO 
\item   to tune  an ensemble learner   to improve
regression and classification performance of models  
\item   
built from the Table~\ref{datadata} data
(and where a  datasets list many goals,
we   predict  for the first listed goal).
\ei
Following this we present results in  Section \ref{results}. See also Section \ref{threats} for a discussion on threats to the validity, and Section \ref{conclusion} for our conclusions.

In summary, the {\bf contributions} of this paper are:
\BLUE
\here{contrib}
\begin{enumerate}
    \item A new pragmatic and  practical  diagnostic for SE optimization: We introduce the Dimensionality Reduction Ratio (DRR) and document a concrete and common, empirical threshold ($DRR > 0.35$), above which optimization becomes remarkably easy.  This threshold acts as a simple, fast diagnostic to determine when complex, expensive optimization methods are unnecessary for a given SE problem.
    
    \item A   correction of a recent TSE paper: We demonstrate that our DRR rule is a more accurate diagnostic than the $I \le 4$ intrinsic dimensionality threshold proposed by Agrawal et al. We show that the $I \le 4$ rule makes the wrong recommendation for most of our datasets, while our rule very often  identifies the "easy" cases.
    
    \item Evidence of a very large  speedup: We show that for the large number of SE tasks with $DRR > 0.35$, a simple optimizer (LITE) achieves statistically identical performance to a complex, state-of-the-art optimizer (DEHB). This allows practitioners to get the same quality of results two orders of magnitude faster (e.g., seconds vs. 20+ minutes), saving significant computational cost.
    
    \item Robust validation of the "DRR Effect": We have observed this effect in  most of the 50 SE data sets studies by this paper.
    
    \item A Full Reproduction Package: We provide a complete and open artifact so other researchers can repeat, refute, or build upon our work. This package includes all 50 datasets, the implementations for both the simple (LITE~\cite{menzies2026aieasylessonslearned}\footnote{\url{https://github.com/timm/ezr}}) and complex (DEHB\footnote{\url{https://github.com/automl/DEHB/}})) optimizers, and the intrinsic dimensionality calculator.
\end{enumerate}
\BLACK

\section{Intrinsic Dimensionality}\label{background}

\BLUE~The assumption   presented in this paper is that, empirically in many SE data we have seen\BLACK:
\begin{itemize}
\item
Problems expressed in $R$ raw attributes can be simplified to
a smaller number of $I$ intrinsic underlying dimensions.
\item 
 Tasks with  $I \ll R$ can be solved   very quickly. We have observed this effect in
 50 of the SE data sets explored here. 
\end{itemize}
This section argues for the first point
(and the rest of the paper explores the second point).

\subsection{Do We Always Need All That Data?}\label{less}
Centuries of research argues that data can be effectively reduced to a smaller set, without introducing errors into the analysis of that data.
In 1901, Pearson argued~\cite{pearson1901principal} that equations involving many variables can often be effectively modeled using fewer variables derived from the eigenvectors of their correlation matrix.
Such a  ``principal component analysis'' can represent many dimensions using just a handful of components~\cite{jolliffe2016principal,xu2019software}.

There is much evidence for this ``reduction'' hypothesis. If a table of data has 
(say) $A=20$ independent   attributes,
and  each attribute
has  $V$ states then for booleans (where $V=2$), that  table needs $V^A=2^{20} > 1,000,000$  rows to cover all possible effects in that space.  But a repeated result for software analytics is that $V^A$  is a massive over-estimation. Models built from $A>20$ attributes can   exhibit high recall, even if that model
is trained from just a few hundred rows~\cite{menzies2008implications}. 

The only way  to explain this is if some small subset $a$   ($|a| \ll |A|$) of the
attributes matter, and the rest can be irrelevant
(e.g. they are either noisy or not associated with the target goal).  Empirically this is indeed the case.
Feature selection research for non-SE data~\cite{hall03,kohavi1997wrappers}  shows that over half the attributes in   tabular data   can be removed, without loss of signal. For SE data, the reduction ratio can be much higher. Pruning   many  rows and columns of   SE datasets often lead to better models~\cite{menzies2021shockingly,rees2017better,Kocaguneli13aa,7194627,chen2005finding}.
The size of the pruning  seen in SE data is   startling. 
For example:
\bi
\item
Chen~\cite{chen2005finding}, Kocaguneli~\cite{Kocaguneli13aa},   Tu~\cite{Tu22},   and Peters \&  Xu et al.~\cite{7194627}
found they could
predict for
Github issue close time,
effort estimation, and 
defect prediction,  even after ignoring      labels  for 80\%, 91\%, 97\%, 98\% (respectively) of their  rows.
 \item
 Later in this paper Figure~\ref{ag1} and Figure~\ref{idvsod} show examples where datasets with dozens to hundreds of attributes can be reduced to half a dozen
 intrinsic dimensions, or less.
\ei
All the above can be summarized as the   ``manifold assumption''~\cite{zhu2005semi}:
\begin{quote}
{\em
Real-world higher-dimensional data often lies on a low-dimensional manifold embedded within the high-dimensional space.}
\end{quote}


\subsection{Calculating Intrinsic Dimensionality}

To find the intrinsic dimensions,
Agrawal et al.~\cite{agrawal2021simpler}   use the 
  fractal-based method of Algorithm~\ref{algo1}.
  In summary, 
this method leverages the concept of a
{\em fractal dimension}, which quantifies the complexity of a dataset by measuring how  things change with the scale of measurement.
More specifically,
 the algorithm checks how many more rows can be found at distance 
$R_{x}$ than  $R_{y}$ for $x< y$ . For example:
\bi 
\item
If the data lies on the ground all over a  football field, then that data lies in a two-dimensional space. Hence,  an increase from $R_{x}$ to  $R_{y}$ will find    {\em polynomially} more examples.
\item
But if the data lies on a straight road that runs through the middle of the field,   then that data is effectively one dimensional and an increase from $R_x$ to $R_y$ will only find {\em linearly} more examples.
\ei
 Algorithm~\ref{algo1} returns the maximum gradient of a curve of $R_x$ vs the log of the number of rows found at distance $R_x$.  
Agrawal et al. report that this algorithm correctly detects up to $I \le 20$  intrinsic dimensions
within   spreadsheets with  low-correlated columns\footnote{To  make low-correlated columns, cells had         random variables.}.

 Algorithm~\ref{algo1}   offers several advantages, including accuracy, robustness, and scalability related to other intrinsic dimensionality calculators. It provides a more accurate estimation of intrinsic dimensions  compared to traditional methods~\cite{camastra2003data} since it is less sensitive to noise and outliers in the dataset. Also, using some stochastic sub-sampling,  it   can be applied to large datasets efficiently. 
Its accuracy, robustness, and scalability make it a valuable tool for dimensionality reduction in various applications~\cite{fragoso2019intrinsic}. 

\begin{algorithm}[!t]
\caption{Calculating intrinsic dimensionality. From~\cite{agrawal2021simpler}.}
\label{algo1}  
\scriptsize\begin{algorithmic}[1]
\State \textbf{Import data from} Testdata.py
\State \textbf{Input:} sample\_num = $n$, sample\_dim = $d$
\State $Rs\_log$ = start : end : step
\State $Rs$ = $\exp(Rs\_log)$
\For{$R$ \textbf{in} $Rs$}
    \State $I = 0$
    \For{$(i, j)$ \textbf{in} combinations(data, 2)}
        \State $d = \text{distance}(i, j)$
        \If{$d < R$}
            \State $I = I + 1$
        \EndIf
    \EndFor
    \State $Cr = \frac{2I}{n(n - 1)}$
    \State $Crs.\text{append}(Cr)$
\EndFor
\For{$i$ \textbf{in} step}
    \State gradient = $\frac{Crs[i] - Crs[i - 1]}{Rs[i] - Rs[i - 1]}$
    \State $GR.\text{append}(\text{gradient})$
\EndFor
\State \text{Smooth}($GR$) \Comment{smooth the curve}
\State $\text{intrinsicD} \gets \max(GR)$ \Comment{return the intrinsic dimensionality}
\end{algorithmic} 
\end{algorithm}

\begin{figure}[!t]
\begin{center}
\includegraphics[width=2.5in]{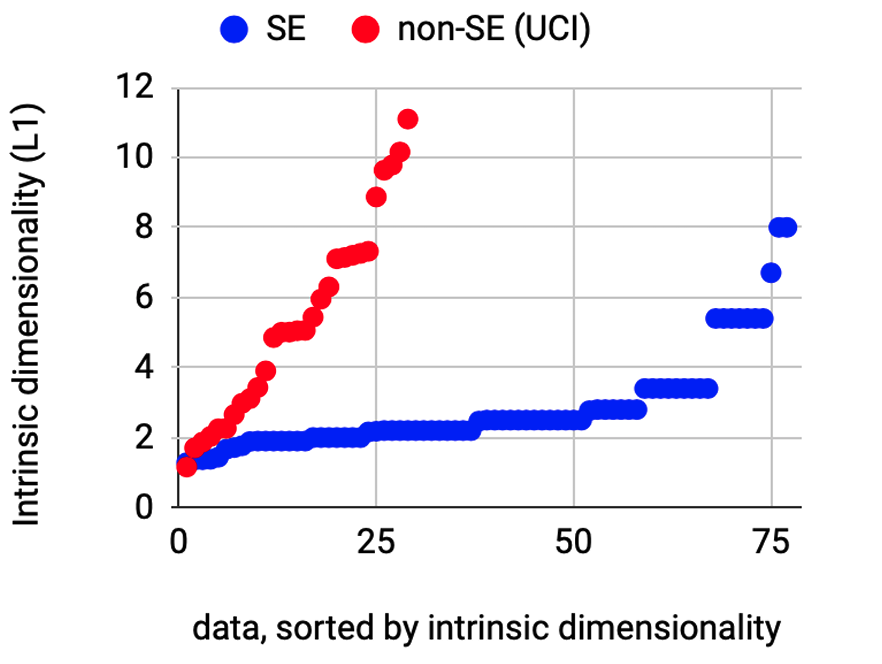}
\end{center}
\caption{Differences in the  intrinsic dimensionality of SE and non-SE data. From ~\cite{agrawal2021simpler}.}\label{ag1}
\end{figure}
\begin{figure}
\begin{center}
\includegraphics[width=3in]{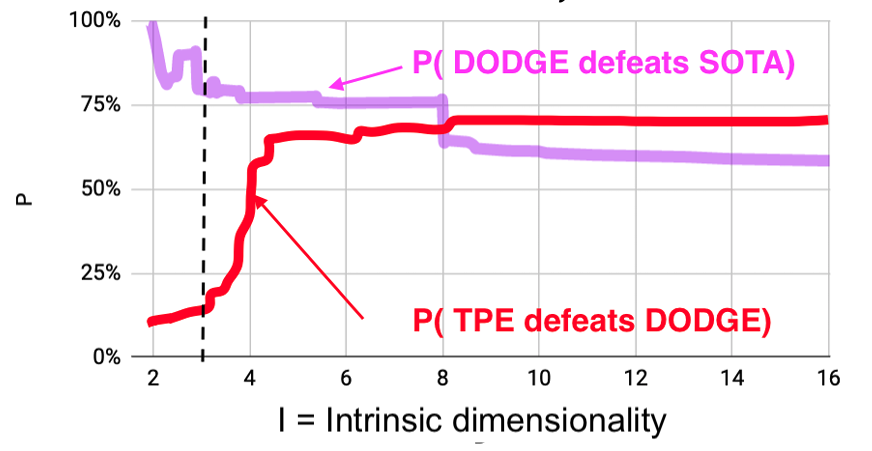}
\end{center}
\caption{According
to Agrawal et al.~\cite{agrawal2021simpler} different algorithms work best at different   intrinsic dimensionalities. Vertical dashed lines shows the median of the SE
data from Figure~\ref{ag1}.}\label{ag2}
\end{figure}

\definecolor{mygreen}{RGB}{0,100,0}
\begin{figure*}[!t]
\begin{center}
\includegraphics[width=\textwidth]{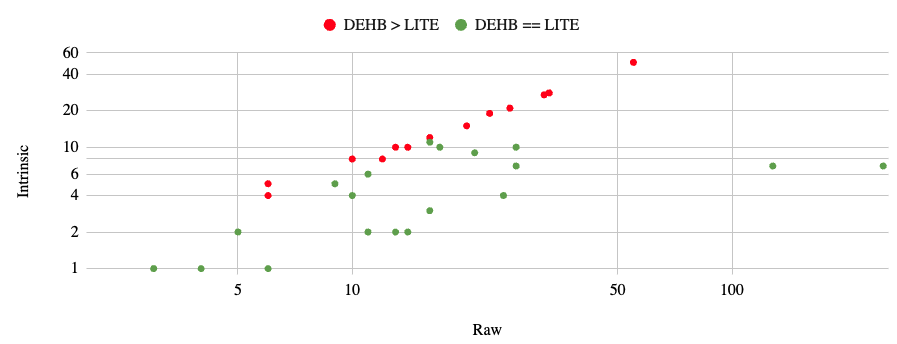}
\end{center}
\caption{Intrinsic Dimensionality vs Original Dimensionality of the Table~\ref{datadata} data.  \textcolor{red}{Red} points indicate data sets where complex optimization (that require 3000 samples) defeated simpler methods (that only required 30 samples).
\textcolor{mygreen}{Green} points are easy cases,
all of which come from SE papers.}
\centering
\label{idvsod}
\end{figure*}
\begin{figure*}
\begin{center}
\includegraphics[width=\textwidth]{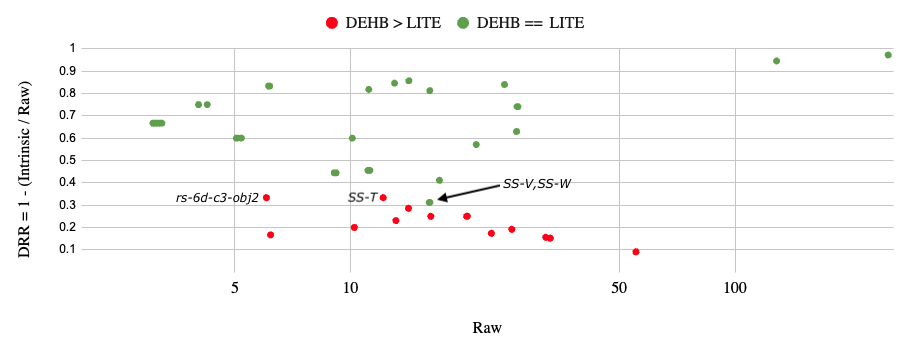}
\end{center}
\caption{Shown here are the Table~\ref{datadata} datasets, scored on the y-axis by Equation~\ref{drr1}.
Colors have the same mean as Figure~\ref{idvsod}.
The   data sets rs-6d-cs-obj2, SS-R, SS-V, SS-W  are discussed in Section~\ref{rq3}.}  
\centering
\label{drralgo}
\end{figure*}

In support of the manifold assumption,
Figure~\ref{ag1} shows 
Agrawal's results for SE and non-SE data.
For SE data  they used data seen for   recent SE analytics 
research publications.
For non-SE data, Agrawal et al.  used datasets from the UCI 
data mining repository~\cite{asuncion2007uci}. This non-SE data has been widely applied in the machine learning community to certify their algorithms~\cite{asuncion2007uci}.

Figure~\ref{ag1} shows that   
Algorithm~\ref{algo1}    reduces data down to   6 intrinsic dimensions (or less).
More importantly, the manifold assumption seems to hold especially true for SE data since, as seen in   Figure~\ref{ag1}:
\bi
\item 
The median SE intrinsic dimensionality is 3.1;
\item 
The median non-SE intrinsic dimensionality is 5;
\item
That is, non-SE  is nearly twice as complex as SE data.
\ei

\subsection{Implications of Low Intrinsic Dimensionality}
Does it matter that SE and non-SE data have different intrinsic dimensionalities?
Agrawal et al. argued this difference  allows them to recommend when to use simpler or more complex hyperparameter optimizers. 
They advocated a simple tabu search optimizer called DODGE\footnote{Given the input settings $S_i$ to a hyperparameter optimizer, and a resulting output performance score $P_i$, then if $P_i$ is similar to an older score $P_j$,  then DODGE deprecated settings near $S_j$.}
 which they compared against a 2012 optimizer called HYPEROPT/TPE (tree of parzen estimators)~\cite{bergstra2012random}.
Their results are shown in 
Figure~\ref{ag2}.
\BLUE\here{clear3}
Agrawal et al. reported that, in their study,
\BLACK
that
different hyperparameter optimizers work best for different intrinsic dimensionalities.
In that figure, for  intrinsic dimensionalities less than 4, their  simpler tabu method was more likely to outperform  TPE in more than half the datasets.   We define      {\em Agrawal's
 threshold} as:  \begin{equation}\label{arule}
I \le 4
\end{equation}

\subsection{Issues with the Agrawal Threshold}\label{bad}
For many reasons, the Agrawal threshold  must now be revisited and revised.

{\bf Firstly}, the 
HYPEROPT/TPE algorithm used in the Agrawal study was proposed in 2012. It is no longer state of the art. This paper must repeat the   Agrawal et al. analysis, but with    more recent algorithms.

{\bf Secondly},
in the Agrawal et al. data sample,
the  right-hand-side of Figure~\ref{ag2} was very   under-populated. This is to say that  their result could have been a conflation of most their data falling to the left of  that figure. This paper will study more data at higher intrinsic dimensionalities.

{\bf Thirdly},
 Figure~\ref{idvsod}
shows the original raw dimensions and the intrinsic dimensionality of the Table~\ref{datadata} data.
In that figure,\BLUE~ the \textcolor{mygreen}{green} points come for optimizing results
where very simple optimizers performed as well (or better)
than very complex optimizers. It should be noted that
for 
those \textcolor{mygreen}  points, simpler optimization was orders of magnitude faster than
more complex methods.
Looking at the locations of the SE and non-SE data, it is clear that
 Agrawal's
 threshold of $I<4$ fails to separate the easy from hard cases.
 On the other hand,
as seen in  Figure~\ref{drralgo},  Equation~\ref{drr1} is   far  better
at distinguishing these two types of data (nearly everything with DRR $> 0.35$ is an ``easy'' case).\BLACK~This separation is important. As shown below,
very different algorithms work best for these separate
groups.

{\bf Fourthly}, and most importantly,
it turns out that Agrawal's rule gives bad advice. As seen in
 Figure~\ref{idvsod}, most of our  data has $I\ge 4$; i.e. which Agrawal would advise ``use a complex optimizer''. Among those datasets, we find many with $\mathit{DRR}>0.35$ which (as shown by the results later in this paper) can be optimized     very simply\footnote{ Specifically,  30 quick samples will result in optimizations
as good as slower methods that require 3000 samples.}.
To say that another, Agrawal's rule is incorrect most of the time.
 
The rest of this paper expands on this last point. 
The next section describes an experiment checking how often
  state of the art AI methods are needlessly computationally expensive in the region   $\mathit{DRR}>0.35$.

\begin{figure*}[!b]
\begin{center}
\includegraphics[width=6in]{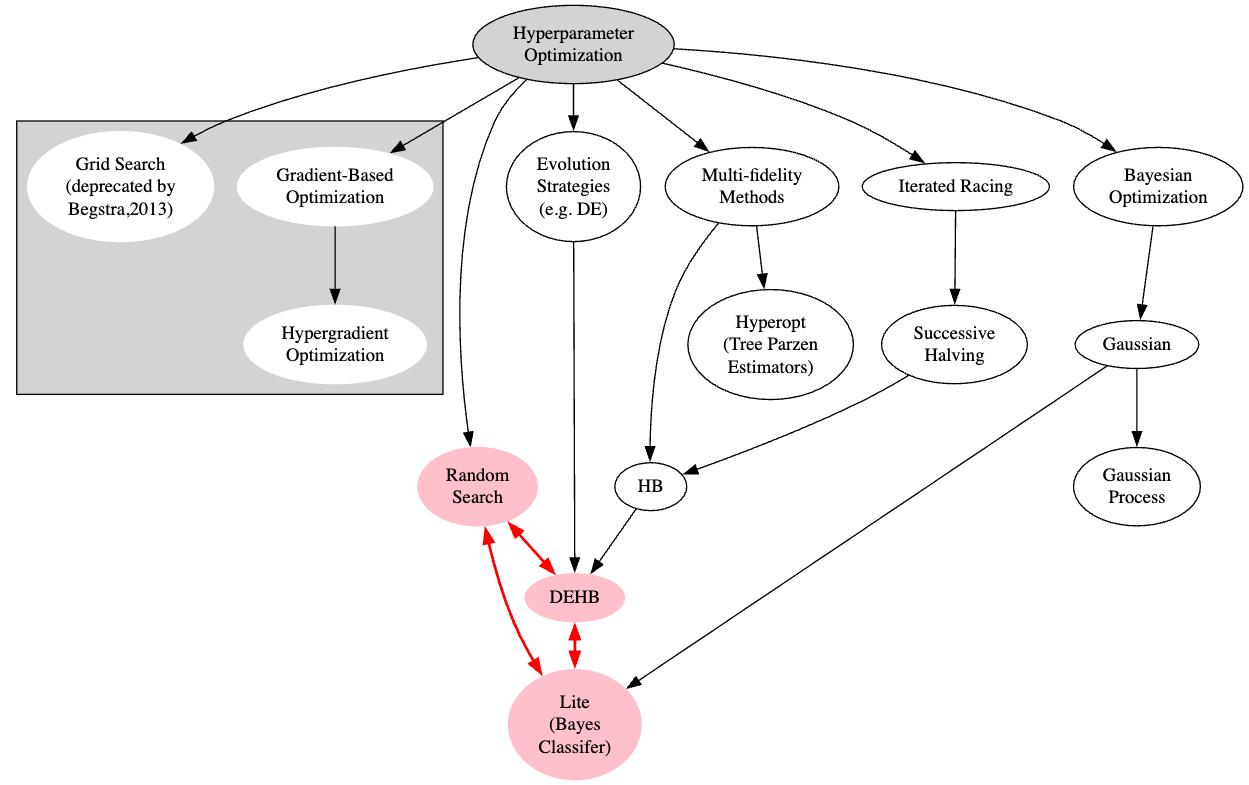}
\end{center}
\caption{A range of hyperparameter optimization methods. From~\cite{bischl2023hyperparameter}. This  paper compares the methods shown in red.  
  }\label{methods1}
\end{figure*}

\section{Methods}
\label{methods}
\subsection{Research Questions}

The methods of this paper aim to answer two questions.
Empirically, for the data studied here:

\bi
\item
  {\bf RQ1:}  Is SE and non-SE data different? 
  \item
  {\bf RQ2:} Is there an effect where  that difference select
  for better HPO? 

  \BLUE 
\ei

\subsection{Data}\label{data}
To answer these questions, we  study   the data shown in Table~\ref{datadata}.
These datasets have a range of independent values  seen in  
Table~\ref{tab:datar}.
The intrinsic dimensionality of those datasets are shown on the x-axis of Figure~\ref{idvsod}.

\BLUE
\here{fam2}
To summarize the kinds of data explored here, we  focus on software engineering problems that can be 
formulated as classification or regression tasks (and the regression can be for $N \ge 1$ goals). Specifically, 
we examine four major families of SE problems:

\textbf{Software Configuration Optimization:} This includes  
software systems where the goal is to predict performance outcomes 
(e.g., runtime, memory usage, energy consumption) or find optimal 
settings. Examples include compiler configurations (LLVM, x264), 
database systems (PostgreSQL, HSQLDB, Redis), web servers (Apache), 
and various software systems (SS-A through SS-X series, representing 
different configuration spaces with 3-88 tunable parameters).

\textbf{Cloud and System Resource Management:} Problems involving 
the optimization of cloud resources and system parameters, including 
Apache server performance tuning, SQL database optimization, video 
encoding settings (X264), and miscellaneous configuration tasks 
across different deployment scenarios.

\textbf{Software Project Health Prediction:} Predicting project 
health metrics and developer activity patterns using historical 
data from repositories. This includes predicting commit rates, 
pull request merge times, and issue closure rates across 35 
different project health datasets.

\textbf{Software Process Models:} Classical SE estimation problems 
including effort prediction, defect forecasting, and schedule 
estimation. This family includes COCOMO-based models (COC1000), 
XOMO variants (Flight, Ground, OSP, OSP2) for different project 
types, POM3 models (A-D) for agile development scenarios balancing 
cost and completion rates, and the NASA93 dataset for multi-objective 
optimization of effort, defects, time, and lines of code.

We acknowledge that software engineering encompasses many other 
important problem types (such as test case generation and 
minimization, program refactoring, requirements prioritization, 
and formal verification) that are not naturally expressed as 
regression or classification tasks. Our work specifically targets 
the subset of SE problems where the goal is to predict numerical 
or categorical outcomes based on historical data or to optimize 
configurations through learned performance models.
\BLACK

\subsection{Algorithms}
\here{lit}
This paper applies {\em hyperparameter optimization}
(HPO) 
to the control parameters of an {\em ensemble learner}
in order to build better predictors
for the dependent attributes in Table~\ref{datadata}.
This section describes ensemble learning,  and the HPO methods 
that might improve them.

\subsubsection{Ensemble Learners}

Ensemble learning trains multiple learners on somewhat different subsets of the data.
This ensemble approach to learning often achieves higher predictive
performance than individual models by aggregating predictions from diverse learners \cite{dietterich2000ensemble} \cite{zhou2012ensemble}. This improvement is particularly significant in complex tasks where single models may struggle.
Also, by combining multiple models, ensemble learning helps mitigate overfitting, leading to better generalization on unseen data \cite{breiman1996bagging} \cite{schapire1999brief}.  Further,
ensemble techniques increase model robustness by reducing the impact of noise and outliers in the data \cite{kuncheva2003measures}. This is useful in applications where data quality may vary.

Random forests are ensembles  
composed of ensembles of multiple decision trees,
each of which is trained on 
a  randomly selected $\sqrt{A}$ sample of the attributes $A$.
Each tree in the ensemble makes a prediction
and the final prediction is some  
aggregation of all the votes.  

This study uses scikit-learn's random forest regressors and random forest classifiers~\cite{scikit-learn}. These make predictions using
the mean and mode (respectively) of their leaf nodes.

To use these random forests, we will ask our HPOs
to make decisions about the following:
\bi
\item
{\em n\_estimators}: number of trees in the forest.
Varied from    1 to 200 (in steps of 10).
\item
{\em criterion} (The {\em error estimator}): when building the tree, how to measure split quality. Options
are
squared\_error, absolute\_error, friedman\_mse,  and poisson.
\item
{\em min\_samples\_leaf}: the minimum number of samples required to be at a leaf node. Varied from  1 to 20.
\item 
{\em min\_impurity\_decrease:}
A secondary stopping criteria used by random forest.
Varies from 1 to 10 (in steps of 0.25)
\item
{\em max\_depth:} 
Max depth of tree. Varied from 1 to 20. 
\ei
In all, this list mentions
$20*4*20*40*20$ = $1,280,000$ different possible configurations.

\subsubsection{Hyperparameter Optimizers}
The goal of hyperparameter optimization
is to  explore tables
 of configuration options
 containing:
 \bi
 \item
 $X$ columns containing configuration settings and 
 \item the $Y$ columns containing the performance scores seen after running one
 configuration option.
 \ei
 We assume the existence of function $F$ such that
 \[Y=F(X)\]
 but we have  no direct access to  that function.

 In this notation, hyperparameter optimization is the selection of some configuration option $c_i$
 from a space of possible configurations $C$.
 These configurations control all the choices within  a learner that returns the model $F$. For example, the end of the last section offered multiple choices for random forests.
 The performance score  $F$  for the predictions made when a learner used
 that configuration.  This
score is computed by observing  the  performance of
\[Y=F(c_i, X)\]


Figure~\ref{methods1} shows a range of HPO approaches. For reasons explained below, we ignore the boxed methods (shown on the left) and focus on the methods highlighted in red.

{\bf Grid Search:}
The slowest way to perform HPO is {\em grid search}, which systematically evaluates every possible configuration. 
  Bergstra et al.~\cite{bergstra2012random} warn strongly against such grid searches.
They note that   different learners/datasets need different grid sizes. 
 A grid small enough to catch all nuances in all learners and datasets would be impractically slow to run. 
In our case, assuming  0.5 seconds to try every configuration (which is our average observed time) and   20 repeated trials (for statistical validity),
then our 1.2 millions options need over 19 weeks of CPU time. 
 All these tests would be independent  so they could  be parallelized to run   in the space of a single eight-hour work day (assuming 25 machines with 16 cores).
 However, and this is our main point, why incur that cost when simpler
 options are available? This is an important question.   When we talk to our industrial colleagues, they point out that reducing the cost of their cloud compute facilities is an increasingly urgent problem.
 For all these reasons, we   seek   alternatives to grid search.

{\bf Gradient-based Optimization:} If exhaustive search is too expensive, a more informed approach might be to study how variables change over time. {\em Gradient-based optimization techniques}, such as those used in neural network training, adjust hyperparameters by following the gradient of the loss function with respect to the hyperparameters~\cite{Maclaurin2015}. These methods are highly effective for continuous hyperparameter spaces and are widely adopted in deep learning. Examples include stochastic gradient descent (SGD) and Adaptive Moment Estimation (the ADAM optimizer)~\cite{Shamir2013,KingmaBa2014}. However, these techniques are not typically used in random forests, which are tree-based, non-parametric methods that do not rely on gradients.

{\bf Random search:} A simple, fast, but potentially incomplete HPO method is {\em random search}~\cite{bergstra2011algorithms}, where $N$ random configurations are sampled and evaluated. Random search has been successfully applied in SE, particularly for optimizing models in defect prediction~\cite{nevendra2022empirical}. 
 Simple random  search is often defeated by  {\em active learning} (discussed below)
that leverages the results of each trial to guide subsequent selections.

{\bf Evolutionary methods:}  Evolutionary methods apply notions from biology to optimization.
{\em Populations} of randomly generated candidates are {\em mutated} (i.e. changed by a small amount). Better candidates are {\em selected} for {\em cross-over}
where multiple candidates are combined to create the next {\em generation}
of candidates.
 Traditional Holland-style  mutation~\cite{holland92} is   computationally expensive since it mutates   $n=100$ candidates 
 for $G=100$ generations (i.e. $10^4$  evaluations in all).
 Storn-style mutation~\cite{storn1997differentialDUPLICATE}, as seen in {\em differential evolution} (DE)
 needs far fewer evaluations. 
DE starts like random search and creates a small initial population of randomly selected configuration options (say, 10 vectors of options per feature being configured). 
\bi
\item
For each generation, each option is  compared to a newly created option that is a mixture of three other options drawn from the population.
The new option replaces the old, when it has a better performance score. 
\item 
After generation \#1, the invariant of DE is that each option in  a  population in generation $g$ is superior to  at least $g-1$ other options. 
Hence, as the algorithm runs, it build new
items   from   increasing  superior options.
\ei
DE  has been widely applied~\cite{pant2020differential}\footnote{
At the core of this writing (Feb 2025), the original DE
paper~\cite{storn1997differentialDUPLICATE} has over 37,000 citations (in Google Scholar).}.
Within software engineering,   DE   has been used for 
optimization tasks such as
 Fu et al.'s tuning study on defect prediction~\cite{fu2016differential};
 Shu et al.'s study on tuning detectors for security issues~\cite{shu2022dazzle},
 and Xia et al's study that tuned project health predictors for open-source JAVA systems~\cite{xia2022predicting}.

 {\bf Multi-Fidelity Methods} do not dictate
 the optimization method {\em per se} but offer a meta-principle for their organization. As such they can be an umbrella technique for controlling other methods. For example
 for neural networks,
  {\em coarse-fine evolution} is a multi-fidelity methods~\cite{pham2018efficient} that applies low-fidelity
 methods   before exploring more expensive ones.
 For another example,
  {\em successive halving} is a 
  multi-fidelity method
  introduced by Li et al.~\cite{li2018hyperband} in
  their 2018 Hyperband (HB) algorithm. This approach employs a novel bandit strategy by simultaneously running multiple configurations with varying resource allocations, such as different numbers of training epochs.
Resources are allocated by progressively eliminating poorly performing configurations. It starts with many candidates, giving each a small budget (e.g., a few training epochs or a dataset subset), evaluates their performance, and discards the worst-performing half at each iteration while doubling the budget for the remaining ones.   

{\bf Iterated Racing:}
This kind of algorithm  iteratively evaluates and compares a set of candidate solutions (or configurations) while discarding under-performing ones.
Successive halving combines multi-fidelity methods with iterated racing.
Other examples of iterated
racing include
the irace method~\cite{lopez2016irace}  of   L{\'o}pez-Ib{\'a}{\~n}ez
et al. Irace implements the
  Iterated F-race algorithm for automatic algorithm configuration
  which iteratively evaluates and refines candidate configurations to efficiently determine optimal parameter settings. Classic 
  iterated racing   reflects on the rankings of different methods.
  Successive halving, on the other hand, also considers the resources required to collect information
  about each configuration.

{\bf Bayesian Optimization:}
 is an  optimization technique used  when
     function evaluations are costly, such as in hyperparameter tuning, experimental design, or engineering simulations~\cite{NIPS2012_05311655}. 
     
     These methods
   build a {\em surrogate model} which can very quickly guess the likely outcome of an evaluation~\cite{deshwal2022bayesian}.
Once a surrogate model is available, the model
built so far can be used to guess what example
should be studied next. This technique is called
{\em active learning}~\cite{settles2009active}.
This is useful since
 when learners choose their own training data, they often build better models with fewer labeled examples~\cite{settles2009active}. 

One common choice for surrogates are  
{\em Gaussian Process Models (GPMs)} which estimate the mean and variance of predictions for unlabeled data.  GPMs fit numerous diverse functions to existing labeled data so their computational cost grows with data size, limiting scalability~\cite{nair2018finding}. 

A faster approach is Tree of Parzen Estimations (TPE) by Bergstra et al.~\cite{bergstra2012random} as implemented in the Hyperopt's Python package\footnote{\url{https://github.com/hyperopt/hyperopt/}}. 
Hyperopt/TPE sorts labeled configurations, splits it at some engineer-defined threshold, then builds surrogate models for the ``best'' and ``rest'' subsets. 
For that modeling, it uses a
Parzen Estimator kernel density estimator.

Faster than Hyperopt/TPE is LITE~\cite{menzies2024streamlining}, a TPE method
that uses    a Bayes classifier to model
  ``best'' and ``rest'':
\begin{enumerate}
\item
Given $M$ evaluated configurations    and
  $N$ unexamined configurations, LITE
  sorts   $M$   into $\sqrt{M}$ ``best'' and 
  the remaining into ``rest'' examples.
  \item
  These   sets   train a two-class classifier that   reports $b,r$; i.e. the likelihood
  of an example being ``best'' or ``rest''.
  \item
 The resulting
  likelihoods are   used by an {\em acquisition function} to select from $N$ the  configuration to run next. 
  \item 
  This results in $M+1$ labeled examples
  and $N-1$ unlabeled examples.  Each new labeling decrements the labeling budget $B$,
  \item While $B>0$, repeat from step \#1.
\end{enumerate}
   On termination, LITE returns the 
  best labeled example seen so far using the measures on Section \ref{eval}.
  For its acquisition function,
  given a   budget of 
  $B$ evaluations   LITE evaluates the configuration from $N$ 
  that maximizes:
  \[
\frac{b+rq}{\mathit{abs}(bq - r + \epsilon)}
  \]
where $\epsilon$ is a very small constant (that avoids
divide-by-zero errors) and 
\begin{equation}\label{q}
q=\begin{cases}
0 \mathit{ \;if \; exploiting}\\
1 \mathit{ \;if \; exploring}\\
1 - \frac{M}{B} \mathit{ \;if \; adapting}
\end{cases}
\end{equation}
Here,  {\em explore, exploit, adapt} are different search strategies for acquiring new information. When labeled examples are  very scarce, it can be best to {\em explore} regions where oracles are     declaring opposite ideas, but with similar weights.
Once some more data has arrived, it can be better to switch to {\em exploiting} that knowledge and just jump to where there are strongest indications of most ``best'' and least ``rest''. Finally, ``adapt'' is a function which, as more $M$    labeled examples 
arrive, the acquisition strategy
slides from {\em explore} to {\em exploit}.

{\bf DEHB:} DEHB extends Hyperband by running multiple sequential halving brackets with different starting budgets, balancing exploration and exploitation. Within each budget level, DEHB uses DE to generate and evolve configurations, maintaining separate subpopulations for each fidelity level. This allows DE to run independently at each budget while enabling information flow from lower to higher budget subpopulations through a modified DE mutation strategy. The first iteration of DEHB resembles vanilla Hyperband, using random search to initialize the lowest fidelity subpopulation. In subsequent iterations, DEHB reuses and evolves subpopulations from previous brackets, eliminating the need for random sampling. 

Our reading of the literature is that, within the conventional AI literature, DEHB~\cite{awad2021dehb}  is arguably the current state-of-the-art
in combining evolutionary methods (using DE's differential evolution), multi-fidelity methods (using HB's Hyperband)  and iterated racing
(using the sequential halving). 
DEHB  out-performs its predecessor (BOHB~\cite{awad2021dehb}) (and BOHB is known  to out-perform Hyperopt~\cite{bergstra2015hyperopt}).

\subsection{Experimental Rig}
To establish   statistical
credence, this rig is run 20 times
with different random seeds.

Each run  optimizes a  random forest    performing regression 
(or classification) for a single dependent variable in each of the datasets. 
The selected dependent  was always the first available in each dataset (and, for this analysis, all other dependents are excluded from the data).

From the algorithms in the last section, we select three for comparison purposes:
\bi
\item DEHB, since it uses a range of state-of-the-art methods (evolutionary methods, multi-fidelity methods, iterated racing).
\item Random search, since it is good  practice to compare stochastic methods with a completely random baseline\footnote{\url{https://github.com/acmsigsoft/EmpiricalStandards/blob/master/docs/standards/OptimizationStudies.md}}.
\item LITE, since, of the above, it is the fastest and uses the fewest samples.
For Equation~\ref{q}, we use $q=1$ since that was recommended by the original LITE paper.
\ei
For random and LITE, we allow up to 30 evaluations.
For DEHB we allow up to  3000 evaluations.
These budgets are those recommended in the LITE
and DEHB papers.

DEHB, with 3000 evaluations, will serve as our exemplar heavyweight method.

LITE, with 30 evaluations, will serve as our   exemplar lightweight method.

For each of these datasets in Table~\ref{datadata},
we started with   10,000 randomly selected 
random forest hyperparameter configurations. 
LITE will   explore up to 30 of
these (selected at random).
DEHB, on the other hand, would sample
100 (at random) then go on to evolve
its own set of preferred configurations.

 \subsection{Evaluation Metrics}\label{eval}

When each algorithm was run on the Table~\ref{datadata} data,
we collected the following evaluation metrics.
We selected these since these are widely seen in the literature\cite{reddy2010software, Sarro16, shepperd2012evaluating,lustosa2024learning}.

Classification and regression problems
need different evaluation criteria. 
For classification,  if $A,B,C,D$ denote
the 
true negatives, false negatives, false positives, and true positives (respectively), then:

\[\begin{array}{rcl}
\mathit{accuracy}&=& (A+D)/(A+B+C+D)\\
\mathit{r\;=\;recall}&=&   D/(B+D)   \\
\mathit{p\;=\;precision}&=& D/(C+D)  \\
\mathit{F1}&=& 2rp / (r+p)\\
\end{array}\]
For regression, 
MRE (magnitude of relative error)  is defined as follows:
\begin{equation}\label{mre}
\mathit{MRE}=
 \mathit{abs}(\mathit{actual}- \mathit{prediction})/\mathit{actual}
\end{equation}
MRE measures how far predicted is from actual.
PRED40, on the other hand, reports on how often a regression prediction falls close the actual. 
PRED40 was recommended by Sarro~\cite{Sarro16}
  and is defined as:
\begin{equation}\label{pred40}
\mathit{Pred40}=
\mathit{Count}(\mathit{MREs}\leq 40 ) / ||\mathit{MREs}||
\end{equation}
Shepperd and MacDonell~\cite{shepperd2012evaluating} argue that regression measures like the above
lack contextual information. Hence they propose 
Standardized Accuracy (SA) which baselines error measures of some algorithm against the errors seen when applying the simplest reasonable estimator (in our case, mean value of the known actuals). SA is based  the Mean Absolute Error (MAE) and is  defined as:
\begin{equation}\label{MAE}
\mathit{MAE} = \frac{1}{N}\sum_{i=1}^{N} |\mathit{prediction}_i - \mathit{actual}_i|
\end{equation}
Here, $N$ is the size of the test set used for evaluating the model performance. SA is then defined as the ratio:

\begin{equation}\label{SA}
\mathit{SA} = (1 - (\mathit{MAE}/\mathit{MAE}_\mathit{dumbPrediction})) \times 100
\end{equation}
and $\mathit{MAE_{dumbPrediction}}$ is the MAE of a set of guesses.

 As per Lustosa et al.'s 2024 TOSEM paper~\cite{lustosa2024learning}, the above    values are combined into a composite metric
{\em d2h} (distance to heaven). To explain {\em d2h}, first we need to explain the  Zitzler    
multi-objective indicator~\cite{zit02}. Zitzler   favors example $B$ over $A$ if jumping from $B$ to $A$ loses more than jumping from $A$ to $B$:

\begin{itemize}
\item Let $ \textit{worse}(A,B)  = \textit{loss}(A,B) > \textit{loss}(B,A)$
\item Let $ \textit{loss}(A,B) = \sum_{j=1}^n -e^{\Delta(j,A,B,n)}/n$
\item Let $ \Delta(j,A,B,n)  =  w_j(o_{j,A}  - o_{j,B})/n$
\end{itemize}
Here $w_i \in \{-1,1\}$ signifies whether we are minimizing or maximizing goal $o_{j}$ (respectively).
\bi
\item
In regression, we seek to    \textit{maximize} Pred40 and SA and \textit{minimize} MRE so $o_j$ values are $\{1,-1,-1\}$ respectively.
\item
In classification, we seek to    \textit{maximize} accuracy, precision,
recall, and F1 so  $o_j$ values are $\{1,1,1,1\}$ respectively.
\ei
Finally in order to summarize all of these metrics into one, {\em d2h} takes all evaluated examples from all techniques and asks how close is each example to the best available in the set.
It can be defined as:
\begin{equation}\label{D2H}
\mathit{D2H_{s}^{i}} = i / |Z|
\end{equation}
where $Z$ is a vector containing all options ranked from best to worst according to Zitzler and $i$ is the index of a sample $s$ in that vector.

\subsection{Statistical Methods}\label{stats}

In our study, we report median and interquartile ranges (which show 50th percentile and 75th-25th percentile), of the D2H metric for each algorithm on each dataset. We collect median and interquartile range values for each of the datasets. 

To make comparisons among all algorithms on a single dataset, we implement the Scott-Knott analysis~\cite{mittas2012ranking}. In summary,
by using Scott-Knott, algorithms are sorted by their performance.
After that, they are assigned to different ranks
if the performance of the algorithm at position $i$ is significantly 
different to the algorithm at position $i-1$.

To be more precise, Scott-Knott sorts the list of experiments (in this paper, LITE, Baseline and DEHB for each of the 16 datasets) by their median score. After the sorting, it then splits the list into two sub-lists. The goal for such a split is to maximize the expected value of differences in the observed performances before and after division~\cite{xia2018hyperparameter}. Scott-Knott analysis then declares one of these divisions to be the best split. The best split should maximize the difference $E(\Delta)$ in the expected mean value before and after the split:
\begin{equation}
    E(\Delta) = \frac{|l_1|}{|l|}abs(\overline{l_1} - \overline{l})^2 + \frac{|l_2|}{|l|}abs(\overline{l_2} - \overline{l})^2
\end{equation}
where:
\begin{itemize}
    \item
    $|l|$, $|l_1|$, and $|l_2|$: Size of list $l$, $l_1$, and $l_2$.
    \item
    $\overline{l}$, $\overline{l_1}$, and $\overline{l_2}$: Mean value of list $l$, $l_1$, and $l_2$.
\end{itemize}
After the best split is declared by the formula above, Scott-Knott then implements some statistical hypothesis tests to check whether the division is useful or not. Here ``useful'' means $l_1$ and $l_2$ differ significantly by applying hypothesis test $H$. If the division is checked as a useful split, the Scott-Knott analysis will then run recursively on each half of the best split until no division can be made. In our study, hypothesis test $H$ is the Cliff's delta non-parametric effect size measure. Cliff's delta quantifies the number of differences between two lists of observations beyond p-values interpolation~\cite{macbeth2011cliff}. The division passes the hypothesis test if it is not a ``small'' effect ($Delta \geq 0.147$).

The cliff's delta non-parametric effect size test explores two lists $A$ and $B$ with size $|A|$ and $|B|$:
\begin{equation}\label{cf}
    Delta = \frac{\sum\limits_{x \in A} \sum\limits_{y \in B} \left\{ \begin{array}{l} +1, \mbox{  if $x > y$}\\
                    -1, \mbox{   if $x < y$}\\
                     0,  \mbox{   if $x = y$}
                \end{array} \right.}{|A||B|}
\end{equation}

In Equation~\ref{cf}, cliff's delta estimates the probability that a value in list $A$ is greater than a value in list $B$, minus the reverse probability~\cite{macbeth2011cliff}. This hypothesis test and its effect size is supported by Hess and Kromery~\cite{hess2004robust}.

\begin{figure*}
\begin{minipage}[!t]{0.60\textwidth}
    \centering
    {\bf Figure~\ref{sedata}.A: SE data sets. D2H median (50th percentile). Lower is better.}
    
    \includegraphics[width=\textwidth]{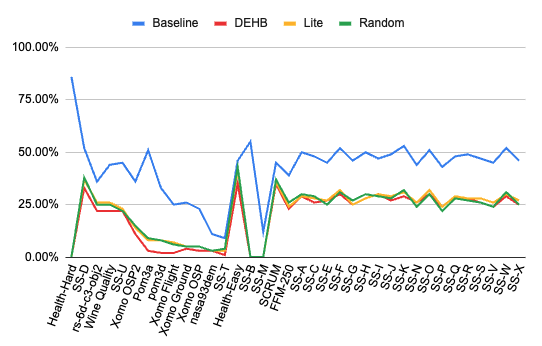}
    
    \vspace{1.8cm}
    
    {\bf Figure~\ref{sedata}.B: SE data sets. D2H IQR ((75th -25th) percentile).\newline Lower is better.}

    ~\\
    
    \includegraphics[width=\textwidth]{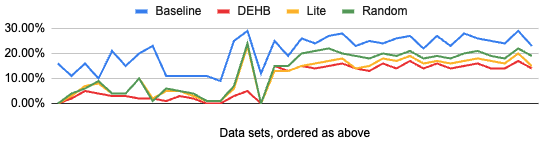}
\end{minipage}%
\hfill
\begin{minipage}{0.38\textwidth}
    \centering
    {\bf Figure~\ref{sedata}.C: mean runtimes (seconds). \newline Lower is better.}
    
    \includegraphics[width=\textwidth]{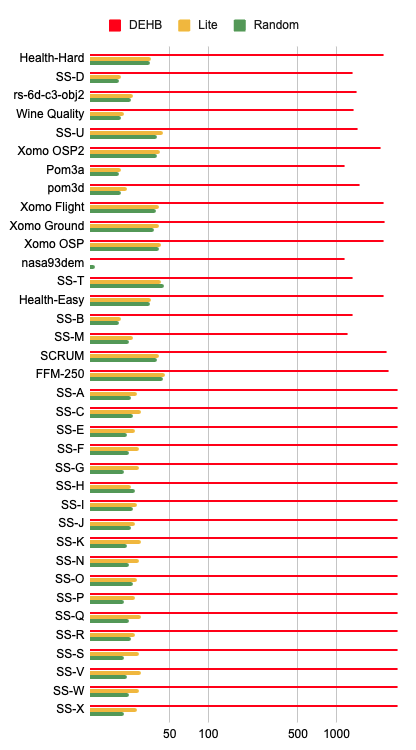}
\end{minipage}
\caption{SE data, results from 20 runs.}
\label{sedata}
\end{figure*}

\begin{figure*}
\begin{minipage}{0.60\textwidth}
    \centering
    {\bf Figure~\ref{uci}.A: Non-SE data.  D2H median (50th percentile).\newline Lower is better.}
    
    \includegraphics[width=\textwidth]{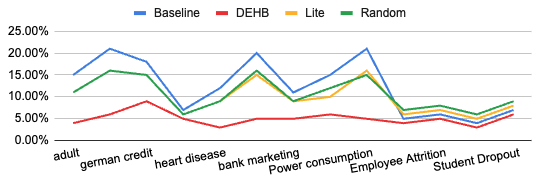}
    
    \vspace{1em}
    
    {\bf Figure~\ref{uci}.B: Non-SE data. D2H IQR ((75th -25th) percentile).\newline Lower is better.}

    ~\\
    
    \includegraphics[width=\textwidth]{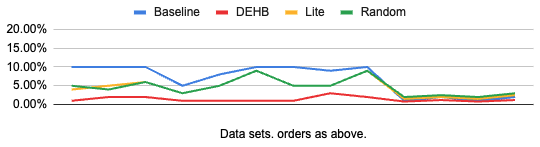}
\end{minipage}%
\hfill
\begin{minipage}{0.38\textwidth}
    \centering
    {\bf Figure~\ref{uci}.C: Non-SE data. mean runtimes (seconds). Lower is better.}
    
    \includegraphics[width=\textwidth]{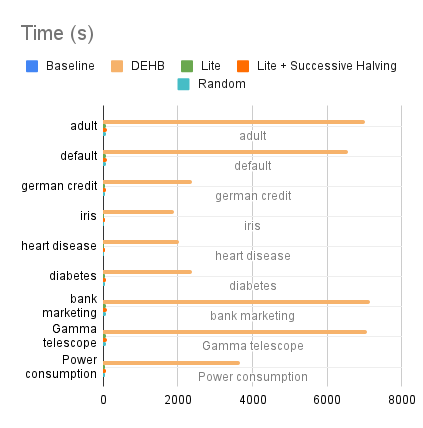}
\end{minipage}
\caption{Non-SE data, results from 20 runs.}
\label{uci}
\end{figure*}

\section{Results}\label{results}

 \subsection{RQ1: is SE and non-SE data different?}
  Prior work, based on the Agrawal threshold 
  Equation~\ref{arule}, argued that this is indeed the case  (see Agrawal et al.~\cite{agrawal2021simpler}). However, that study had preponderance of low dimensional data. Recall from Section~\ref{bad} that when  we revisited that conclusion, for the data studied here,
  Figure~\ref{idvsod} showed many cases where Equation~\ref{arule}
  would incorrectly select for ``hard'' datasets  half the time ($\frac{34}{48}=71\%$). Hence, we cannot   endorse Equation~\ref{arule}.

  That said, when we (a)~extend the data analysis to higher dimensional data; and (b)~replace Equation~\ref{arule} with Equation~\ref{drr1},
  we find ourselves in general agreement with Agrawal.
  Figure~\ref{idvsod} shows that,
 for at least in the sample tabular data studied here:
 \begin{formal}
{\bf Answer1:}  SE data usually has higher DRRs (i.e. a lower ratio of intrinsic dimensions) than    non-SE data.
\end{formal}

\BLUE
\subsection{RQ2: Does that difference select for better HPO?}\label{rq3}
\here{rq3}

To answer this question, we applied simple and complex optimizers to a large
sample of SE problems and non-SE problems, then checked if the DRR threshold
\[
 \left(1-I/R\right) \ge 0.33
\]
separates the problems that simple methods can solve.

For the non-SE data, like Agrawal et al., we use the  UCI 
data mining repository~\cite{asuncion2007uci}.

For the SE data, we turned to the MOOT repository of multi-objective
optimization tasks~\cite{Moot:2025}. Curated by   Menzies and   Chen, this
repository contains dozens of multi-objective optimization problems collected
from recent papers in top SE venues such as the International Conference on
Software Engineering~\cite{chen2026promisetune}, Foundations of SE (FSE)
conference~\cite{nair2017using}, IEEE Trans. SE~\cite{chen2025accuracy}, the
Information Software Technology journal~\cite{CHEN2018281}, Empirical Softw.
Eng.~\cite{peng2023veer}, Mining Software Repositories~\cite{nair18}, IEEE
Access~\cite{lustossa2024isneak}, ACM Trans. SE
Methodologies~\cite{lustosa2024learning} and the Automated Software Engineering
Journal~\cite{nair2018faster}.
For this paper, we picked   50 projects at random from MOOT.

Turning now to the results, Figure~\ref{sedata}.A shows median {\em D2H} for
SE data:
\bi
\item
The upper curve shows the baseline of the untreated data (expressed as median
{\em D2H} scores).
\item Below that, the other curves fall on top of each other.
\item
After applying the statistics of Section~\ref{stats}, we can confirm the
visual impression that the performance of these lower curves is
indistinguishable.
\item That is, for this data, a state-of-the-art AI optimizer (DEHB) that
needs 3000 evaluations does no better than a very lightweight optimizer (LITE)
using 30 evaluations.
\ei
On the other hand, Figure~\ref{uci}.A shows non-SE data:
\bi
\item
As before, the upper curve shows the baseline (median {\em D2H}) of the
untreated data).
\item Below that, we see other curves with more visual separation than before.
\item
The treatment with the lowest curve (i.e. closest to heaven) is the treatment
that uses 3000 evaluations (DEHB).
\item That is, for this data, it is not enough to merely do 30 evaluations.
\ei
Figure~\ref{sedata}.B and Figure~\ref{uci}.B show that none of these methods,
other than the baseline and random selection, have a stability problem. Which
means that both DEHB and LITE are stable for these datasets (low variance in
repeated trials). Figure~\ref{sedata}.C and Figure~\ref{uci}.C comment on the
runtime cost of running 30 versus 3000 evaluations. In both curves, the LITE
method runs two orders of magnitude faster (a few seconds as opposed to 20
minutes, or more).

So when does 30 evaluations suffice? Figure~\ref{drralgo} sorts all 50
projects by DRR. Recall that most of the SE data
($\frac{38}{50}=76\%$) have high DRR values. In Figure~\ref{drralgo}, the DRR
threshold seen in our original 24 projects moved very slightly, from 0.33 to
\[
 \left(1-I/R\right) > 0.35
\]
(since, above this threshold, 100\% of the data sets are easy to optimize). We
characterize this as a very small change between the original report (24
projects) and the larger sample (50 projects).

As mentioned in the introduction, the DRR effect is an empirical effect and
not some binding law of the universe. Hence, there are some exceptions to this
effect. Consider in Figure~\ref{drralgo}:
\begin{itemize}
\item SS-V, SS-W, which have a DRR of 0.30;
\item rs-6d-c2-obj2 and SS-T, which have a DRR of 0.33.
\end{itemize}
Note that the latter two data sets, with the larger DRR, are harder to
optimize. Given the DRR effect, this is the opposite of what we would expect.
Clearly, around the threshold where the effect takes hold, there is some
noise. Accordingly, we set the DRR threshold to above 0.35.

Hence we say:
\begin{formal}
{\bf Answer2:} {\here{1.1}}~By separating data on DRR, we can find problems
that can be solved effectively, two orders of magnitude faster (than using
state-of-the-art AI HPO algorithms). While our sample of 50 projects does not
guarantee universality over all SE projects, it does show that the DRR effect
appears frequently. Practitioners should check for high DRR before deploying
expensive optimization methods. This simple diagnostic could save orders of
magnitude in computational cost.
\end{formal}
\BLACK

\section{Discussion}

\subsection{Details}
When discussing this work with colleagues, we are often asked the following questions.

{\bf   Can Equation~\ref{drr1}
 simplify all SE analytics?} 
No.
Generation problems need the complexities of different classes of algorithms. Also,
the certification requirements of safety-critical
software is not a simple process.
That said,
 we do show here that  better and faster results can be
obtained selecting algorithms via intrinsic problem complexity. 
This is an important message since, to our shame,
  SE researchers rarely benchmark complex   methods against simpler approaches\footnote{
For example, in  a recent  systematic review~\cite{Hou24} of 229 SE papers using large language models (LLMs),  only $13/229 \approx 5\%$ of those papers compared LLMs to other approaches even though  other methods can
 produce results that are better and/or 
faster~\cite{grinsztajn2022why,somvanshi2024survey,Tawosi23,majumder2018500+,Ling24,Fu17easy,johnson2024ai}.}.

{\bf    Why does DRR matter?}   DRR divides data into:
\bi
\item
One group where many attributes have to be explored;
\item And   another group
where many attributes 
are superflous and can be ignored.
\ei
Complex AI algorithms are needed to explore datasets where most of the attributes are important. Otherwise, in SE data,  simpler and faster methods may suffice.

{\bf  Why   so many superfluous attributes in SE data?}
Software construction is a complex combination of  tasks. Data collected from software projects is  hence   a report of many things,
not all of which are   relevant to particular goals. 
Hence,  we should expect that many attributes from SE data can be ignored.

{\bf    Why has Equation~\ref{drr1}
  not been previously reported in the AI literature?}  
Most of our SE datasets ($\frac{38}{50}=76\%$) have high DRRs 
but most of our AI datasets have much lower
DRRs.  Hence, AI researchers   have  missed our result since they were studying different  data.

This observation has a methodological implication. A common practice in SE analytics is to use  AI algorithms off-the-shelf.
The results of this paper suggest that this practice needs to be deprecated.
Other researchers agree:
\bi
\item 
Novielli et al. report that
sentiment analysis tools perform much better for SE problems when they are specifically trained on SE data~\cite{10.1145/3196398.3196403}.
\item
Binkley et al.~\cite{binkley2018need}
warn that off-the-shelf
information retrieval tools
need to be significantly
adjusted before they are applied to SE applications.
\ei
{\bf  When Equation~\ref{drr1} holds, we can recommend something simpler than LITE?} \label{faq4}
Figures~\ref{sedata}.A and ~\ref{sedata}.B showcase that random selection (of 30 examples) performs just as well as LITE on the SE data. Nevertheless, we would still endorse LITE over random.
Firstly, while random is simpler than LITE, LITE is hardly a complex algorithm 
(evidence: see its 30 line implementation \footnote{
\url{https://github.com/timm/ezr/blob/main/ezr.py\#L518C1-L548C68}}).
Secondly, it we look into how random
would be applied in practice,
then ``random'' starts looking a lot like
LITE:
\bi
\item
If new data arises, random search would require 30 more labels to handle that data while LITE could just reuse its existing model (i.e. zero extra labels).
\item
On the other hand, the  random selection results from the old data  could be used to build a classifier (for best and rest) and that classifier could be used (without further labels) to predict for the new data. 
\item
However, if anyone asks for validation results from that classifier, statistics would have to be collected over the old data. If it was suggested to collect those statistics incrementally during label collection over the old data, then this   random-plus-classifier approach 
would be almost the same as LITE.
\ei





\BLUE

\definecolor{LightGray}{HTML}{BDBDBD}
\begin{table*}[t]
 
\centering
\caption{Positioning DRR relative to prior and emerging work.}
\label{tab:related}
\scriptsize
\begin{tabular}{p{2cm}p{3.8cm}p{5.2cm}p{5.4cm}}
 
\textbf{Category} &
\textbf{Representative Works} &
\textbf{Core Idea} &
\textbf{Relation to DRR} \\ \midrule
Hyperparameter Optimization &
Agrawal et al.\,\cite{agrawal2021simpler};
Van Aken et al.\,\cite{vanaken17} &
Automate tuning of learner parameters to maximize performance. &
DRR identifies when simple HPO suffices, avoiding unnecessary complexity. \\ \arrayrulecolor{LightGray}\hline
Reduce Features or Dimensions &
Kohavi \& John\,\cite{kohavi1997wrappers};
Jolliffe\,\cite{jolliffe2016principal} &
Reduce irrelevant or redundant features to improve generalization and runtime efficiency. &
DRR quantifies intrinsic vs.\ observed dimensions, extending classical feature analysis to SE data. \\\hline
Empirical SE\newline Effects &
Boehm\,\cite{boehm1981software};
Brooks\,\cite{brooks1975mythical};
Glass\,\cite{glass2002facts};
Wohlin \& Runeson et al.\,\cite{wohlin2012experimentation} &
Recurring empirical regularities that guide SE practice (e.g., Brooks’ Law, Pareto principle). &
DRR acts as a similar empirical effect linking data complexity to optimization difficulty. \\\hline
LLM Input \newline Simplification &
DietCode\,\cite{zhang22m};
SlimCode\,\cite{wang24slime} &
Prune or compress source-code tokens to lower inference cost and preserve model accuracy. &
Conceptually parallel: both seek efficiency by removing redundant information before learning. \\\hline
LLM Architecture Simplification &
Avatar\,\cite{shi24avatar};
SimPy\,\cite{shi24avatar} &
Compress or redesign model grammars and architectures to reduce compute, energy, and tokenization overhead. &
Complementary: these simplify the \emph{model};
DRR simplifies the \emph{problem representation}. \\ 
 \arrayrulecolor{black}
 
\end{tabular}
\end{table*}

\subsection{Related Work}

\here{related}
Prior studies connect to this paper through three themes: (a) tuning
and optimization, (b) dimensionality reduction, and (c) empirical
effects guiding SE practice.  Table~\ref{tab:related} summarizes key
parallels.  We also note a recent trend on simplifying large language
models (LLMs) for code, which shares the same philosophy of matching
solution complexity to problem complexity.

The first group of releated work concerns \textit{hyperparameter optimization (HPO)}.
Earlier work such as Agrawal \textit{et al.}\,\cite{agrawal2021simpler}
and Van Aken \textit{et al.}\,\cite{vanaken17} explored
automatic tuning of learner or system parameters.  DRR differs by
measuring \emph{intrinsic problem complexity} to decide when simple
optimizers are sufficient.

The second group involves \textit{feature selection and
dimensionality reduction}, including the classical wrappers of
Kohavi and John\,\cite{kohavi1997wrappers} and principal-component
methods summarized by Jolliffe\,\cite{jolliffe2016principal}.  DRR extends
these ideas to modern SE datasets, quantifying how many underlying
dimensions truly drive behavior.

The third line treats SE ``laws'' and \textit{empirical effects}
as practical heuristics for reasoning about projects
\cite{boehm1981software,brooks1975mythical,glass2002facts,wohlin2012experimentation}.
Our work follows this tradition, proposing the DRR effect as an
empirically recurring pattern that links data structure to optimizer
choice.

Finally, several recent papers explore how large language models (LLMs)
for code can be simplified without losing accuracy.  Works such as
DietCode\,\cite{zhang22m} and SlimCode\,\cite{wang24slime}
focus on the \emph{input side} of model efficiency.  DietCode prunes
low-information tokens and statements based on attention weights from
pre-trained models, showing that 40\% of the code can be removed with
minimal loss of performance.  SlimCode generalizes this idea, introducing
a model-agnostic approach that relies on structural properties of source
code rather than model attention, achieving comparable accuracy while
reducing training time and inference cost by more than an order of
magnitude.  Together these studies show that simpler representations can
make even large models faster and cheaper to deploy.

Complementary work targets the \emph{model architecture itself}.
Avatar from Shi et al.\,\cite{shi24avatar} formulates the reduction of model size,
inference latency, energy consumption, and carbon footprint as a
multi-objective optimization problem solved by a satisfiability modulo
theory (SMT)–based tuner.  The resulting models are up to 160$\times$
smaller and 184$\times$ more energy efficient than their original
counterparts.  SimPy\,\cite{shi24avatar} takes a different direction by
redesigning the Python grammar for AI consumption, removing redundant
formatting tokens, shortens keywords, and proposes a dual grammar where
humans use Python while models use a condensed variant.  Both methods
explicitly link computational cost to representational redundancy, an
idea that resonates with DRR.

Across these LLM efforts, the unifying principle is that performance
and efficiency improve when representation complexity is matched to the
task’s intrinsic structure.  In that sense, DRR and LLM simplification
studies belong to the same methodological family: DRR measures
redundancy at the \emph{data level} to decide when simpler optimizers
suffice, while the LLM community removes redundancy at the
\emph{representation or model level} to make heavy architectures more
tractable.  Both lines of work reflect a broader trend toward empirical
diagnostics that quantify when ``less'' can in fact be ``better.'' 

Given the similarities in DRR, DietCode, SlimCode, Simpy and Avatar we conjecture
that there is so as-yet-unrealized fusion of research combining tabular data landscape methods
(e.g. DRR) with LLMs.  We have preliminary ideas on this, but nothing definitive to report at this time. 

\BLACK

\subsection{Threats to Validity}\label{threats}

 As with any empirical study, different biases can threaten the final results. Therefore, any conclusions born from this work must be considered with the following concerns in mind.

\noindent \textbf{Parameter Bias:}
In terms of parameter bias, the LITE algorithm uses one of many proposed acquisition functions towards active learning. There are other options of acquisition functions that may still be explored and that may outperform this current version of LITE. DEHB was also ran with default settings as suggested in their previous work \cite{awad2021dehb}, there may yet be other settings that can perform better than LITE in these problems.

\BLUE

\noindent \textbf{Sampling Bias:}
\here{sampling}
We have selected 50 datasets to represent various families of SE problems. 
These datasets may not be representative all SE problems.
As discussed above, the effect we report has survived a  "doubling study" (where we examined twice the number of data sets). Still, that coverage does not encompass the entire range seen in software engineering. 
We only have two data sets with very large attribute sets (SCRUM has 128 attributes and FFM-250 has 250 attributes).
Notably, these datasets with a large number of raw attributes showed few intrinsic dimensions (for these data sets, both had $I=7$), raising questions about what is the  range of intrinsic dimensions  for real-world data.  Future work should explore the typical intrinsic dimensions of real-world datasets.

\BLACK


\noindent \textbf{Algorithm Bias:}
We have only selected DEHB as a baseline, our reasoning behind this is the absolute supremacy this algorithm achieved in the field of hyperparameter optimization when compared to the remaining state-of-the-art algorithms. There are multiple studies that compare DEHB to other state-of-the-art algorithms or use DEHB in problems similar to those tackled here. However, this does not guarantee that there is not a better suited algorithm that could outperform both LITE and DEHB in the case studies provided in this work.

\subsection{Future Work}

Our findings open several avenues for further research.

\BLUE
First , expanding our dataset collection to include industrial SE projects would help validate whether the observed trends hold beyond controlled benchmarks. 
This paper has explored the optimization of classification and  regression for 50 SE datasets. Future work should explore more data and more tasks.
\BLACK

Second, integrating DRR estimation into automated optimization pipelines could lead to adaptive algorithm selection in real-time.

Thirdly, refining the DRR  measurement techniques could improve the accuracy of complexity estimation, reducing reliance on fixed thresholds like Agrawal’s (Equation~\ref{arule}). 

Fourthly, and perhaps more importantly,
we have shown that under   certain circumstances (see Equation~\ref{drr1}),
some SE tasks can be solved very simply. Hence it may be useful to revisit decades of SE analytics to find those tasks that seemed to be very slow, but can now be solved very simply.

\section{Conclusion}\label{conclusion}

This study challenges the conventional wisdom that complex AI-based optimizers are always necessary for software engineering (SE) problems. By analyzing intrinsic dimensionality (ID) and dimensionality reduction ratios (DRR), we have shown that many SE datasets contain redundant features that allow for simpler, faster optimization methods. Our results suggest that instead of defaulting to computationally expensive techniques, researchers and practitioners should first assess dataset complexity using DRR before selecting an optimization approach.

Through extensive experiments across multiple SE and AI datasets, we find that traditional hyperparameter optimization strategies can often be replaced by lightweight heuristics like LITE and Successive Halving without loss of effectiveness. 

\BLUE
\here{conc} As stated in the introduction, we  stress that the DRR effect
is a common
empirical effect that holds for 50 datasets from a range of SE tasks (see Section~\ref{data}).
As such it is a {\em limited conclusion} and   not some ironclad natural law (as evidence of this, in Section~\ref{rq3} we discuss four counter-examples to this effect).

Nevertheless, the DRR effect is frequently found in our data.
 We hence recommend that practitioners check
for high DRR before deploying expensive optimization
methods. As said above,  the DRR effect is a  simple diagnostic that could save
orders of magnitude in computational cost.
\BLACK


\section*{Conflict of Interest Statement}
The authors declared that they have no conflict of interest.
Permission is  granted, free of charge, to any person obtaining a copy
of this software and associated documentation files (the ``Software''), to deal
in the Software without restriction, including without limitation the rights
to use, copy, modify, merge, publish, distribute, sublicense, and/or sell
copies of the Software, and to permit persons to whom the Software is
furnished to do so, subject to the   
conditions of the MIT License.

{\footnotesize \bibliographystyle{IEEEtran}
\bibliography{main}
}

\balance

\begin{IEEEbiography}[{\includegraphics[width=1in,height=1.25in,clip,keepaspectratio]{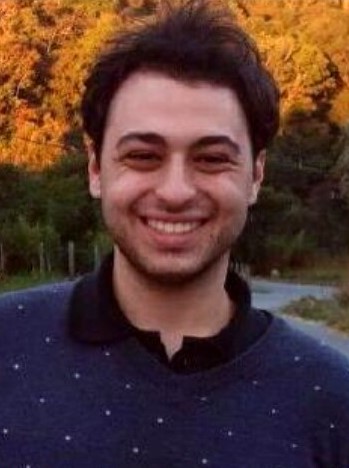}}]{Andre Lustosa}
received his PhD in Computer Science from North Carolina State University in 2025. He is a Principal Software Engineer
  and Team Lead at Red Hat, where he leads efforts to build and package the AI kernel across Red Hat's AI product
  portfolio. His research interests include software engineering, optimization, and data mining. He is an active open
  source contributor and recipient of the Red Hat AI Jedi Award (Q4 2025). For more information, please visit  \url{http://alustos.us}.
  \end{IEEEbiography}

\begin{IEEEbiography}[{\includegraphics[width=1in,clip,keepaspectratio]{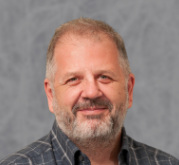}}]{Tim Menzies} (ACM Fellow, IEEE Fellow, ASE Fellow, Ph.D., UNSW, 1995) is a full Professor in Computer Science at North Carolina State. He is the director of the   Irrational Research Lab (mad scientists r'us) and the author of over 300 publications (refereed) with 24,000 citations and an h-index of 74. He has graduated 22 Ph.D. students, and has been a lead researcher on projects for NSF, NIJ, DoD, NASA, USDA  and private companoes (total funding of \$19+ million). Prof. Menzies is the editor-in-chief of the Automated Software Engineering journal and associate editor of TSE and other leading SE journals. For more, see  \url{https://timm.fyi}.
\end{IEEEbiography}

\newpage

\clearpage


\end{document}